\documentclass[12pt, draftclsnofoot, onecolumn]{IEEEtran}
\usepackage{amssymb}
\usepackage{amsmath}
\usepackage{cite}
\usepackage{url}
\usepackage{xcolor}
\usepackage{cite,graphicx,amsmath,amssymb}
\usepackage{subfigure}
\usepackage{citesort}
\usepackage{fancyhdr}
\usepackage{mdwmath}
\usepackage{mdwtab}
\usepackage{amsthm}
\usepackage{setspace}

\usepackage{yhmath}
\usepackage{bm}
\usepackage{comment}

\newtheorem{remark}{Remark}
\newtheorem{theorem}{Theorem}

\newtheorem{lemma}{Lemma}

\newtheorem{corollary}{Corollary}
\newenvironment{corollarybox} {\begin{corollary}}{\hfill \interlinepenalty500 $\Box$\end{corollary}}

\begin{document}

\title{A Novel Physics-based Channel Model for Reconfigurable Intelligent Surface-assisted Multi-user Communication Systems}

\author{

% Yuanwei\ Liu, Zhijin\ Qin, Maged\ Elkashlan, Yue\ Gao, and Lajos\ Hanzo
Jiaqi~Xu and
Yuanwei~Liu,~\IEEEmembership{Senior Member,~IEEE}

\thanks{J. Xu and Y. Liu,  with Queen Mary University of London, London,
UK (email:\{jiaqi.xu, yuanwei.liu\}@qmul.ac.uk).}
}

%\author{
%\IEEEauthorblockN{  Jiaqi~Xu\IEEEauthorrefmark{1}, and Yuanwei~Liu\IEEEauthorrefmark{1}} \\
%\IEEEauthorblockA{
%\IEEEauthorrefmark{1} Queen Mary University of London, London, UK\\
%\IEEEauthorrefmark{2} King's College London, London, UK\\
%\{jiaqi.xu, yuanwei.liu\}@qmul.ac.uk\\
%\ arumugam.nallanathan@kcl.ac.uk\\
%}}
%\author{
%\IEEEauthorblockN{ Yuanwei~Liu\IEEEauthorrefmark{1}, Zhijin~Qin\IEEEauthorrefmark{1}, Maged Elkashlan\IEEEauthorrefmark{1}, and  Yue~Gao\IEEEauthorrefmark{1}\\} \IEEEauthorblockA{
%\IEEEauthorrefmark{1} Queen Mary University of London, London, UK\\
%%\IEEEauthorrefmark{2} Lancaster University, Lancaster, UK\\
% } }

%\author{

% Yuanwei\ Liu, Zhijin\ Qin, Maged\ Elkashlan, Yue\ Gao, and Lajos\ Hanzo

%Yuanwei~Liu,~\IEEEmembership{Student Member,~IEEE,}
%        Zhijin~Qin,~\IEEEmembership{Student Member,~IEEE,}
%        Maged~Elkashlan,~\IEEEmembership{Member,~IEEE,}
%        Yue~Gao,~\IEEEmembership{Senior Member,~IEEE,}
%        and Lajos~Hanzo,~\IEEEmembership{Fellow,~IEEE}

%\thanks{Y. Liu, Z. Qin, M. Elkashlan, and Y. Gao are with Queen Mary University of London, London,
%UK (email:\{yuanwei.liu, z.qin, maged.elkashlan, yue.gao\}@qmul.ac.uk).}
%\thanks{L. Hanzo is with University of Southampton, Southampton,
%UK (email:lh@ecs.soton.ac.uk).}
%\thanks{ Z. Ding is with Lancaster University, Lancaster, UK (e-mail: z.ding@lancaster.ac.uk).}
%}

\maketitle
\begin{abstract}
The reconfigurable intelligent surface (RIS) is one of the promising technologies contributing to the next generation smart radio environment. A novel physics-based RIS channel model is proposed. Particularly, we consider the RIS and the scattering environment as a whole by studying the signal's multipath propagation, as well as the radiation pattern of the RIS. The model suggests that the RIS-assisted wireless channel can be approximated by a Rician distribution. Analytical expressions are derived for the shape factor and the scale factor of the distribution. For the case of continuous phase shifts, the distribution depends on the number of elements of the RIS and the observing direction of the receiver. For the case of continuous phase shifts, the distribution further depends on the quantization level of the RIS phase error. The scaling law of the average received power is obtained from the scale factor of the distribution. For the application scenarios where RIS functions as an anomalous reflector, we investigate the performance of single RIS-assisted multiple access networks for time-division multiple access (TDMA), frequency-division multiple access (FDMA) and non-orthogonal multiple access (NOMA). Closed-form expressions for the outage probability of the proposed channel model are derived. It is proved that a constant diversity order exists, which is independent of the number of RIS elements. Simulation results are presented to conﬁrm that the proposed model applies effectively to the phased-array implemented RISs.
\end{abstract}

\begin{IEEEkeywords}
Channel model, non-orthogonal multiple access, phase errors, reconfigurable intelligent surface, reflect-array, Rician fading.
\end{IEEEkeywords}
\section{Introduction}
The reconfigurable intelligent surface (RIS), also known as intelligent reflecting surfaces (IRS), is a two-dimensional (2D) material structure that is reconfigurable in terms of its electromagnetic wave response \cite{di2020smart}. These 2D surface structures can be implemented by metamaterials \cite{chen2016review}, phased-array antennas \cite{arun2020rfocus}, or other technologies. Among these different RIS implementations, a major class of RISs can be modeled, base on local design, as periodic unit cells integrated on a substrate. For RIS implemented by phased-array antennas, the cells do not interact with each other. The electromagnetic wave response of the RIS, such as phase discontinuity, can be reconfigured by tuning the surface impedance, using various mechanisms. Apart from electrical voltage, other mechanisms are reported, including thermal excitation, optical pump, and physical stretching. 
\begin{comment}
Existing designs of patch-array RIS achieve discrete phase control or amplitude control: Arun~\emph{et al.}~\cite{arun2020rfocus} designed the \textit{RFocus}, which is a two-dimensional surface with a rectangular array of passive antennas. Each passive unit is $\lambda/4 \times \lambda/10$ of size and either let the signal go through or reflect it. The authors show that the \textit{RFocus} surface can be manufactured as an inexpensive thin wallpaper, requiring no wiring, and improves the median signal strength by $9.5$ times. Welkie~\emph{et al.}~\cite{welkie2017programmable} developed a low-cost device embedded in the walls of a building to passively reflect or actively transmit radio signals. Dunna~\emph{et al.}~\cite{dunna2020scattermimo} present \textit{ScatterMIMO}, which uses a smart surface to increase the scattering in the environment. In their hardware design, each reflector unit uses a patch antenna which further connects to 4 open-ended transmission lines. The transmission lines provide $0$, $\pi/2$, $\pi$ or $3\pi/2$ phase shifts. Moreover, each unit has its addressable memory. The \textit{ScatterMIMO} achieves a throughput gain of $2$ times and a MIMO SNR impartment of $4.5$dB compared to the baseline. 
\end{comment}
Although this periodic structure has been intensely studied in the field of applied physics and antenna theories, analytical derivations are still required before they can be reconciled to the channel models in communication theories. The concentrations of the modelling work of the RIS-assisted channel are on the links associated with each of the distinct elements on the RIS. There are two types of links, the links from the transmitter to each RIS elements, and the links from RIS elements to the receiver. In the following paper, we use the term half-channels to refer to both of these links. We use the term linked-half-channels to refer to the transmitter-$n$-th element-receiver link. We use the term joint channel to refer to the overall transmitter-RIS-receiver channel. 

\subsection{Prior Works}
Existing research contributions model the RIS channel through optimization perspective \cite{wu2019beamforming,wu2019intelligent,mu2020joint} or through performance analysis perspective \cite{basar2019wireless,badiu2019communication,zhang2020reconfigurable,zhang2019analysis,ding2020impact}. 
In~\cite{wu2019beamforming} and \cite{wu2019intelligent}, the Wu~\emph{et al.} derived the power scaling law of the Rayleigh distributed half-channels when the direct link between the transmitter and the receiver is ignored. The results showed that for the case of random phase shift, the average received power scales with the number of elements ($N$), for the case of optimal phase shift, the power scales with $N^2$. In~\cite{mu2020joint}, Mu~\emph{et al.} developed suboptimal algorithms for RIS with different multiple access schemes. The half-channels were assumed to have Rician distributions. In~\cite{basar2019wireless}, Basar~\emph{et al.} showed that the maximized signal-to-noise ratio of the RIS-assisted channel follows a non-central chi-square distribution, under the assumptions that the half-channels follow Rayleigh distributions. In~\cite{badiu2019communication}, Badiu~\emph{et al.} studied the impact of RIS phase noise for Rayleigh and Rician distributed half-channels. In~\cite{zhang2020reconfigurable}, Zhang~\emph{et al.} modelled each linked-half-channels as a Rician distributed channel. A power scaling law similar to the one given in~\cite{wu2019beamforming} was obtained. 

Next, we focus on four important subjects related to RIS channel modelling:
\subsubsection{Path Loss Model}
Representative works on the path loss model of the RIS channel are as follows: In~\cite{tang2019wireless}, three path loss formulas were proposed for far-field beamforming case, near-field beamforming case, and near-field broadcasting case. For the near-field broadcasting case, the path loss is proportional to $(d_1+d_2)^2$, where $d_1$ and $d_2$ denotes the distance from the transmitter to the RIS and from the RIS to the receiver, respectively. For other cases, the path loss is proportional to $(d_1d_2)^2$. 
In~\cite{ozdogan2019intelligent}, {\"O}zdogan~\emph{et al.} further disproved the $(d_1+d_2)^2$ formula for the far-field case and present the path loss model at an arbitrary observation angle.
\subsubsection{Multipath Fading Model}
Representative works on the multipath fading (small-scale fading) characterize each half-channel from the transmitter to the $n$-th element on the RIS and from the $n$-th element to the receiver by well-known distribution, such as Rayleigh fading~\cite{ding2020impact} and Rician fading~\cite{zhang2019analysis}. The overall multipath fading channel was the multiplication of the above two types of links and the phase shift matrix. In~\cite{badiu2019communication} and \cite{qian2020beamforming}, Badiu~\emph{et al.} and Qian~\emph{et al.} considered the fading channel with the presence of phase errors and its influence on the signal-to-noise ratio.

\subsubsection{Typical RIS Functions}
In the literature, the RIS functions under these two working scenarios were often referred to as \textit{anomalous reflection} and \textit{beamforming}~\cite{liu2020reconfigurable}. Anomalous reflection is a wavefront transformation from a plane wave to another plane wave, while beamforming is a wavefront transformation from a plane wave to a desirable wavefront. For the case where the half-channels are line-of-sight links, the optimal RIS phase configurations for these two functions are governed by two different principles: the generalized laws of refraction and reflection~\cite{bell1969generalized} for \textit{anomalous reflection}, and the co-phase condition~\cite{huang2008reflectarray} for \textit{beamforming}.

\subsubsection{Multiple Access in RIS-assisted Networks}
Research contributions of applying orthogonal multiple access (OMA) and non-orthogonal multiple access (NOMA) in RIS-assisted networks includes \cite{ding2020simple,fu2019intelligent,mu2019exploiting,zheng2020intelligent}. In~\cite{ding2020simple}, Ding~\emph{et al.} proposed an RIS-assisted NOMA transmission architecture where the RIS only serves cell-edge users. In~\cite{fu2019intelligent}, Fu~\emph{et al.} investigated the joint beamforming design of the downlink multiple-input single-output (MISO) RIS-assisted NOMA networks. In~\cite{mu2019exploiting}, Mu~\emph{et al.} optimized the sum rate of MISO IRS-NOMA networks. In~\cite{zheng2020intelligent}, Zheng~\emph{et al.} compared the performance of NOMA and OMA in RIS-assisted networks for different user pairing strategies.

\subsection{Motivation and Contribution}
We observe the fact that previous research contributions studied the path loss effect and multipath fading effect separately. Moreover, research contributions for the RIS hardware capabilities, the physics models for the RIS radiation patterns, and channel models in communication theory need to be reconciled. Under typical RIS working conditions where the RIS function as an anomalous reflector, the joint channel can be modelled more compactly. As a result, we motivate our work as follows:
\begin{itemize}
    \item For the line-of-sight (LoS) dominate link, RIS typically functions as an anomalous reflector. In these cases, separately model the half-channel result in a loss of physical and geometrical information, since the phases of the adjacent half-channels are correlated.
    \item For the application scenario where the RIS-assisted link varies over time, in-time channel estimation for each half-channel proves to be difficult, even impossible~\cite{bjornson2020reconfigurable}. Models need to be proposed for performance analysis where the RIS configures itself base on the information of the joint channel, instead of a collection of $n$ half-channels.
    \item The physical parameters of the system, including phase error caused by the quantization level of the RIS, the geometrical size of each RIS element (compared with the wavelength), the direction of the receiver all have different effects on the overall RIS-assisted channel. For an optimal RIS configuration, these effect needs to be characterized in the channel distribution in closed-forms.
\end{itemize}

Motivated by the above challenges, we aim to propose a novel joint channel fading model for a typical RIS application scenario. Consider a transmitter locating in an environment which is not rich in local scattering, so that the transmitted plane wave can arrive at the RIS with a small AOA spread. Moreover, by appropriately placing the RIS, the receiver has an LoS link with the RIS, but not with the transmitter. Under this setting, the joint channel from the transmitter to the receiver contains a specular component that is dominated by the RIS LoS link and a scatter component contributed by the non-line-of-sight (NLoS) direct link. Statistical channel analysis is carried out by considering $M$ RIS multipath components and $N$ multipath components for the scattering environment. The primary contributions of this paper are as follows:
\begin{itemize}
    \item We propose a novel model in which the performance of the RIS-assisted wireless channel is investigated. The critical methodology is the use of radiation pattern calculation, combined with statistical multipath analysis. In our analysis, we first clarify the adopted RIS hardware model and the communication signal model. Then, we propose both the path loss model and the multipath fading model for the joint channel. 
    \item For the case where RIS functions as an anomalous reflector, we derive closed-form expressions for the joint channel distribution, considering both continuous phase shifts and discrete phase shifts. Base on the derived distribution, we further analyze the outage probability, as well as the scaling law of the average received power. 
    \item We evaluate the performance of RIS-assisted networks where multiple users are served by one RIS. Different multiple access schemes are compared, including time-division multiple access (TDMA), frequency-division multiple access (FDMA), and NOMA. We demonstrate that for each user, a constant diversity order exists, which is independent of the number of RIS elements. The derived analytical results are further confirmed by Monte Carlo simulations.
\end{itemize}

\subsection{Organization}
The rest of the paper is organized as follows. 
In Section \ref{sys_model}, we propose our channel model for the RIS-assisted joint channel, which includes hardware model, signal model, path loss model, and multipath fading model. Section \ref{case} presents closed-form channel distribution for the case where the RIS works as an anomalous reflector. In Section \ref{analysis}, we study the RIS-assisted channel with different multiple access schemes and their outage probabilities. Specifically, the case where multiple users served by a single RIS is studied. Numerical results are presented in Section \ref{num} to verify our analysis, which is followed by our conclusions in Section \ref{conclusion}.

\section{Channel Model and Radiation Calculation}\label{sys_model}
\subsection{Hardware Model: Phased Array-based RIS}\label{pre}

The electromagnetic characteristics of the RIS, such as phase discontinuity, can be reconfigured by tuning the surface impedance. Various mechanisms support this tuning. Apart from electrical voltage, other mechanisms are reported, including thermal excitation, optical pump, and physical stretching. Among them, electrical control is the most convenient choice, since the electrical voltage is easier to be quantized by field-programmable gate array (FPGA) chips. The choices of materials of the RIS include semiconductors~\cite{zhu2013active} and graphene~\cite{emani2015graphene}. Regardless of the different tuning mechanisms, the general geometry layout of the phased array-based RIS can be modelled as periodic unit cells integrated on a substrate. When designing RIS-assisted communication systems, the most important parameter of the RIS is the reflection coefficient $\Tilde{r}$ at each element (cell). To characterize the tunability of the RIS, the method of equivalent lumped-element circuits can be adopted.
\begin{figure}[t!]
\centering
\includegraphics[width =3.0in]{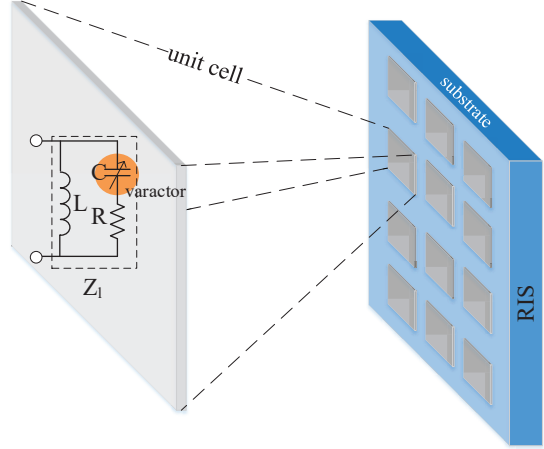}
\caption{Schemetic diagram of the varactor RIS}
\label{vara}
\end{figure}
As shown in Fig. \ref{vara}, the unit cell is equivalent to a lumped-element circuit with a load impedance of $Z_l$. Particularly, the equivalent load impedance could be tuned by changing the bias voltage of the varactor diode. According to the boundary conditions of electromagnetic field (EM) fields, the reflection and transmission coefficient is determined by the impedance $Z_l$ and vacuum impedance $Z_0 \approx 377\Omega$. For the case of normal incidence, it can be shown that:
\begin{equation} \label{im}
    \Tilde{r} =\frac{\bm{E_r}}{\bm{E_i}} = \frac{Z_l-Z_0}{Z_l+Z_0},
\end{equation}
where $\Tilde{r}$ is the complex reflection coefficient, $\bm{E_r}$ and $\bm{E_i}$ are the reflected electric field and the incident electric field. In \eqref{im}, we present the relation between the phase discontinuity $\phi$ and equivalent surface impedance $Z_l$. Discrete phase shift control can be realized by tuning $Z_l$ upon adding different levels of bias voltages. When modelling the RIS in wireless communication system designs, we can characterize each of its unit cell by the local reflection coefficients. For example, the $i$-th cell can be modelled as:
\begin{equation}\label{refl}
    \Tilde{r_i}=\beta_i \cdot e^{j\phi_i},
\end{equation}
where $\beta_i$ and $\phi_i$ correspond to amplitude control and phase control, respectively. In the following sections, $\phi(m,n)$\footnote{It is worth pointing out that the phase discontinuity denoted by $\phi$ should not be confused with other notations denoting geometrical angles, such as $\theta$,$\varphi$ or $\alpha$} refers to the phase discontinuity of the $(m,n)$-th element on the RIS, $\Tilde{r}(x,y)$ refers to the reflection coefficient as a function of the position on the RIS plane.

\subsection{Signal Model}
\begin{figure}[t!]
\begin{center}
\subfigure[Network model]{
        \includegraphics[width=2.7in]{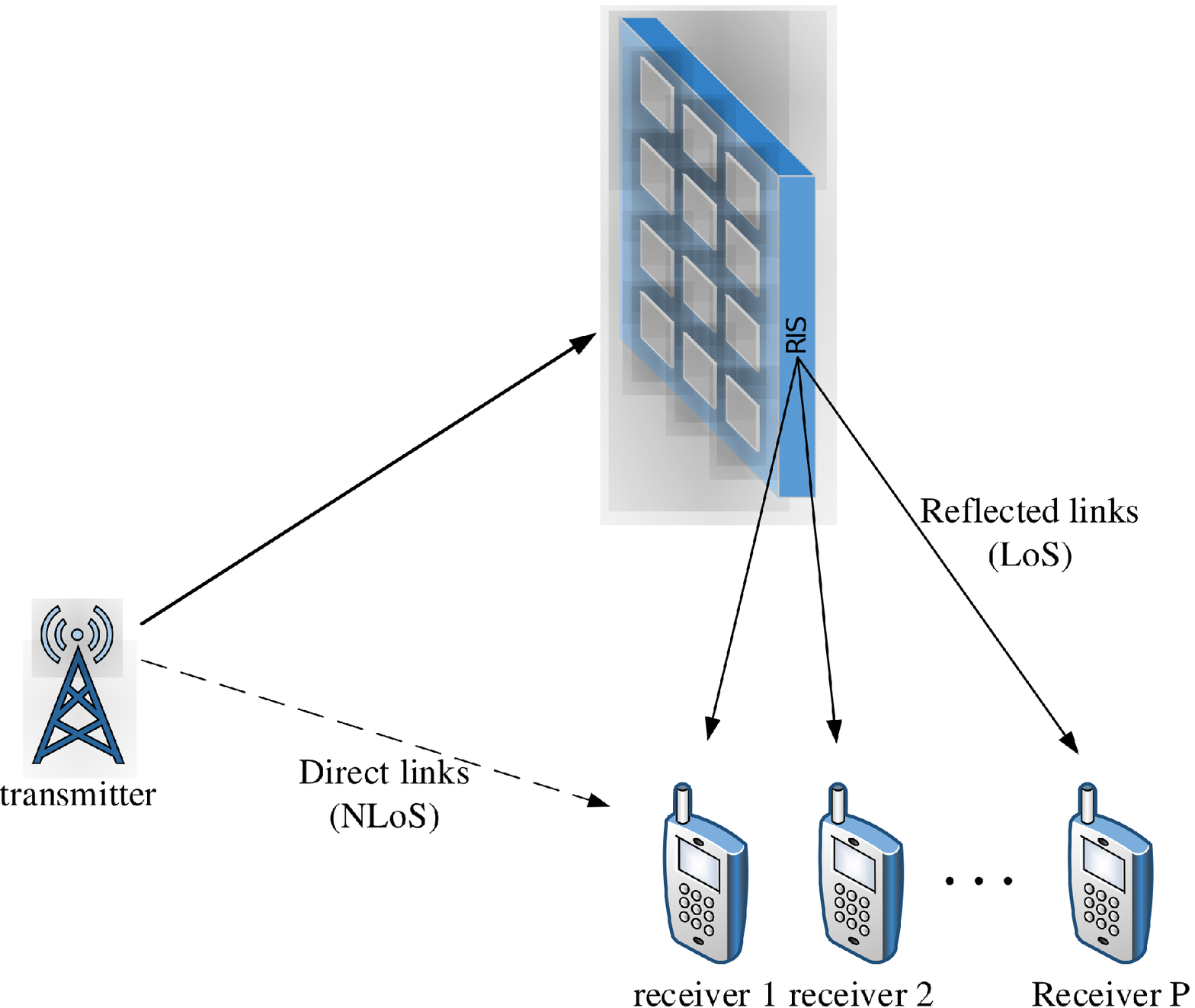}
        \label{sys}
}
\subfigure[2-D model for the RIS link]{

        \includegraphics[width=2.7in]{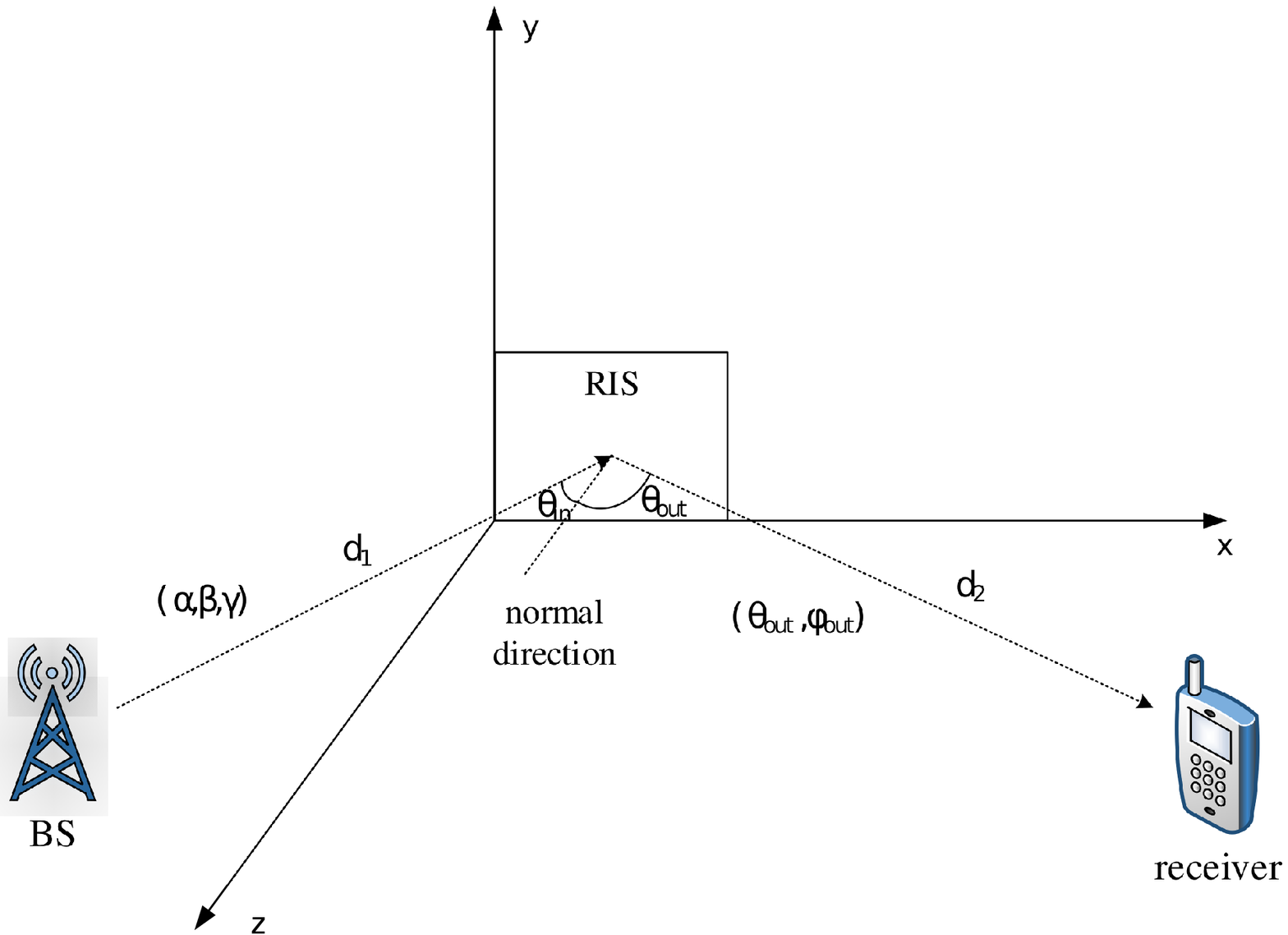}
        \label{2d_sys}
}
\end{center}
\caption{Illustrations of the system model}
\end{figure}

In the following, a novel channel model for the RIS assisted wireless communication system is studied. Consider a wireless channel between a transmitter and $p$ receivers through an RIS. Moreover, we assume the channel between the transmitter and the receivers are flat fading channels. As a result, we can characterize the channels by studying the unmodulated carrier signal. Assume the transmitted signal is of the form:
\begin{equation}\label{trans}
    s(t)=Re[e^{j\frac{2\pi}{\lambda_c}t}].
\end{equation}
The passband signal received by each receiver can be expressed in the quadrature form
\begin{equation}\label{rec}
    r(t)=T_c(t)\cos(\omega_ct)-T_s(t)\sin(\omega_ct),
\end{equation}
where $T_c(t)$ and $T_s(t)$ are the inphase and quadrature components. For convenience, we define $\tilde{R}(t)$ to be the complex envelope of $r(t)$:
\begin{equation}\label{complex}
\tilde{R}(t) = T_c(t)+jT_s(t),
\end{equation} 
where $j$ denote the imaginary unit. As illustrated in Fig.\ref{sys}, consider two links between the transmitter and each receiver: the direct link, which is an NLoS link, and the specular link through the RIS, which is an LoS link. Moreover, we denote the number of RIS columns and rows by $M_x$ and $M_y$, and the number of multipath components considered in the direct link by $N$. 
\begin{theorem}

\label{tt1}
For the RIS-assisted fading channel, assuming vertical polarization for the wireless signal, the complex envelope $\tilde{R}(t)$ have the following form:
\begin{equation}\label{rec_2}
	\tilde{R}(t)= \tilde{R}^{RIS}(t) + \tilde{R}^{D}(t),
\end{equation}
where
\begin{equation}\label{r_ris}
\tilde{R}^{RIS}(t,\theta_{out},\varphi_{out})=IDFT2[c(m,n)e^{\omega_0t+\theta_0+\phi(m,n)}],\\
\end{equation}
\begin{align}\label{1idft2}
\begin{split}
IDFT2&[F(m,n)](p,q)=\frac{1}{M_xM_y}\cdot \\
&\sum_{m=0}^{M_x-1} \sum_{n=0}^{M_y-1}F(m,n)e^{j\frac{2\pi m}{M}p}e^{j\frac{2\pi n}{N}q},\\
\end{split}
\end{align}
\begin{equation}
\tilde{R}^{D}(t)=\sum_{n=1}^{N}b_n e^{\omega_nt+\theta_n},
\label{r_d}
\end{equation}
where $(\theta_{out},\varphi_{out})$ are the angles through which the receiver observes the signal (as shown in Fig.~\ref{2d_sys}). $c(m,n)$ are the amplitude of the signal reflected through the $(m,n)$-th element of the RIS, $\omega_0=\frac{2\pi v}{\lambda_c}\cos(\gamma-\alpha_0)$, $v$ is the velocity of the receiver, and $\gamma$ denotes the angle of this movement. $\theta_0$ is a fixed phase shift angle once the geometry of the system is fixed.
$\phi(m,n)$ is the additional phase shift by the $(m,n)$-th element of the RIS, $b_n$ are the amplitude of the $n$-th multi-path signal, $\omega_n=\frac{2\pi v}{\lambda_c}\cos(\gamma-\alpha_n)$, where $\alpha_n$ is the angle of arrival w.r.t x-axis of this n-th multi-path signal, and $\theta_n = 2\pi (f_c+f_{D,n}\frac{L_n}{c})$, where $f_{D,n}=\frac{v}{\lambda_c}\cos(\alpha_n)$ is the Doppler shift frequency of the n-th multi-path signal, $L_n$ is the total distance for the n-th multi-path signal to travel and $c$ is the speed of light. $p,q$ are related to the observing angle of each receiver.
\begin{proof}
\textit{The proof and detailed expressions for $\theta_0$ and $c(m,n)$ can be found in Appendix A.}
\end{proof}
\begin{remark}
According to \eqref{r_ris}, the received envelope of the specular components through the RIS is a function of the observing angles: $(\theta_{out},\varphi_{out})$. The derivations in Appendix A suggests that the relationship between the indexes $(p,q)$ and $(\theta_{out},\varphi_{out})$ can be explicitly written as:
\begin{align}\label{expl}
    &\sin\theta_{out} \cos\varphi_{out}=\frac{2\pi}{M_xp_xk_0}p, \\ 
&\sin\theta_{out} \sin\varphi_{out}=\frac{2\pi}{M_yp_yk_0}q. 
\label{exp2}
\end{align}
When applying IDFT2, indexes $(p,q)$ can only take on $M_x \times M_y$ different integer values, and as a result, $\tilde{R}^{RIS}(\theta_{out},\varphi_{out})$ can be evaluated in $M_x \times M_y$ distinct directions. However, one can compute $\tilde{R}^{RIS}(\theta_{out},\varphi_{out})$ at more directions by extending the grid on the RIS plane and setting the amplitude equal to zero for all the elements outside of the RIS.
\end{remark}

\end{theorem}
\textbf{Theorem \ref{tt1}} states that the overall received complex envelope consists of two parts, representing the specular link through the RIS and the direct link, respectively. The direct link is composed of a number of $N$ multipath components each with a magnitude of $b_n$ and a phase delay of $\theta_n$. The specular link through the RIS is written as a 2-D inverse discrete Fourier transform (IDFT2) of the reflected EM field patterns at the RIS. The IDFT2 is a double summation over each row and column of elements on the RIS. This result is sensitive to the change of observing angle ($\theta_{out},\varphi_{out}$). 

Next, we consider the far-field scenario where the change in height is negligible compared with the horizontal distance travelled by the signal. As illustrated in \ref{2d_sys}, the system is a 2-D problem where the transmitter, the RIS, and the receiver are located in the $y=0$ plane. In this case, we have $\varphi_{in}=\varphi_{out}=0$, so that $q=0$ in \eqref{exp2}. As a result, the RIS only steers the beam within the $x-z$ plane and we have the bellow corollary: 

\begin{corollary}\label{c1}
For the 2-D problem, the inphase and quadrature components of the received envelope $\tilde{R}(t)$ have the following form:
\begin{equation}\label{tc}
	T_c(t)=\sum_{m=1}^{M}c_{m}\cos(\omega_0t+\theta_0-\epsilon m+\phi_{m}) + \sum_{n=1}^{N}b_n\cos(\omega_nt+\theta_n),
\end{equation}
\begin{equation}\label{ts}
	T_s(t)=\sum_{m=1}^{M}c_{m}\sin(\omega_0t+\theta_0-\epsilon m+\phi_{m}) + \sum_{n=1}^{N}b_n\sin(\omega_nt+\theta_n),
\end{equation}
where $c_{m}$ are the amplitude of the signal reflected through the $m$-th column on the RIS. $\epsilon = 2\pi \frac{p_x}{\lambda_c}(\sin\theta_{out}-\sin\theta_{in})$, $p_x$ is the period length of the elements along $x$ direction of the RIS, $\phi_{m}$ is the additional phase shift by the RIS $m$-th column.
\begin{proof}
\textit{The proof can be found in Appendix B.}
\end{proof}
\end{corollary}

In the following discussions, we focus on the 2-D problem and present our path loss model and multipath fading (small-scale fading) model.

\subsection{Path Loss Model}
According to the analytical results presented in Appendix B, the amplitude of each multipath components of the specular link through the RIS (the linked-half-channel), namely $c_m$, is proportional to the element size ($p_xp_y$), the inverse of the multiplied distance ($1/(d_1d_2)$), the Fraunhofer diffraction factor ($sinc(\frac{ku'p_x}{2})$), and the leaning factor ($\cos\theta_{out}$). These are the large scale attenuations that can be analysed as the path loss. For the case where the RIS is located in the far-field of the transmitter, without preforming amplitude adjustments, we have: $c_m=c_0= \sqrt{P_{spec}}\cdot PL(d_1, d_2, \theta_{in}, \theta_{out}), \forall m \in (0,M]$, where $PL(d_1, d_2, \theta_{in}, \theta_{out})$ denotes the path loss of the joint specular link through the RIS.
\subsection{Multipath Fading Model}

The distribution of the joint channel squared envelope, namely $|\tilde{R}(t)|^2$, is of particular interest. First, we give the general methodology to study the multipath fading of the joint channel. In Section \ref{case}, for scenarios where the RIS function as an anomalous reflector, we derive the closed-form expressions for the joint channel distribution.
\subsubsection{Methodology}
According to \textbf{Theorem \ref{tt1}} and \eqref{complex}, the real (inphase) and imaginary (quadrature) part of the overall envelope both have two terms:
\begin{align}\label{terms}
&T_c(t)=Re[\tilde{R}(t)]=Re[\tilde{R}^{RIS}]+Re[\tilde{R}^{D}],\\
&T_s(t)=Im[\tilde{R}(t)]=Im[\tilde{R}^{RIS}]+Im[\tilde{R}^{D}].
\end{align}
To characterize the overall by a Rician distribution, we need to calculate the mean value for both $T_c(t)$ and $T_s(t)$, as well as their variance. Then, each part can be approximate by a Gaussian process with a non-zero mean. Finally, base on their derived first and second-order moments, we can approximate the magnitude of the joint channel using well-known distributions, such as Rician distribution or Nakagami's m-distribution~\cite{nakagami1960m}.
\subsubsection{Mean value}
It can be proved that the direct link component does not contribute to the non-zero mean value. As a result, the specular link component needs to be analyzed. For the case where $c(m,n)$ is constant for each element, at the same time, RIS phase shifts are perfectly accurate so that $\Phi(m,n)$ does not distribute in a certain error range, the mean value of the overall envelope is equal to its specular component, i.e. $\mathbb{E}[\tilde{R}(t)]=\tilde{R}^{RIS}(t)$. For more general case where both $c(m,n)$ and $\Phi(m,n)$ are characterized by different distributions, the mean value of $T_c$ and $T_s$ needs to be calculated base on these distributions.

\subsubsection{Variance}
Since the specular link and the direct link components are not correlated, their variances can be calculated separately. For the direct link part, it can be proved that $Var(|\tilde{R}^D(t)|)=N\cdot\mathbb{E}[b_n]$. For the specular part, when $c(m,n)$ and $\Phi(m,n)$ do not exhibit any distribution, $\tilde{R}^{RIS}(t)$ is fixed for any given time $t$, so that the variance is zero. For other cases, the variance of $\tilde{R}^{RIS}(t)$ needs to be further calculated.

\subsubsection{Channel Distribution}
Base on the above calculations, we can approximate the joint channel distribution using Rician distribution or Nakagami's m-distribution. Suppose we have:
\begin{align}\label{moments}
&\Xi^2=\mathbb{E}^2[T_c]+\mathbb{E}^2[T_s],\\
&\sigma=Var(T_c)+Var(T_s)=2\cdot Var(T_c).
\label{var}
\end{align}
\begin{theorem}
\label{tt2}

If the variances of $T_c$ and $T_s$ are the same, as indicated in \eqref{var}. Moreover, if the covariance between $T_c$ and $T_s$ is zero, the magnitude of the complex envelope $|\tilde{R}|$ can be approximate by Rician distribution with the effective shape factor:
\begin{equation}
K^{Eff}=\frac{\Xi^2}{\sigma},
\end{equation}
or by a Nakagami m-distribution:
\begin{equation}\label{naka_m}
m = \frac{(\sigma^2+\Xi^2)^2}{(\sigma^2+\Xi^2)^2-\Xi^4}.
\end{equation}
\begin{proof}
According to \eqref{tc} and \eqref{ts}, if we considered a number of multipath that is the same as the number of RIS columns, i.e. $M=N$. Then, both $T_c(t)$ and $T_s(t)$ can be seen as sums of $M$ independent random variables which are drawn from two fixed probability distributions. When $M$ is sufficiently large, according to central limit theorem, $T_c(t)$ and $T_s(t)$ can be treated as random Gaussian process with non-zero means. As a result, the magnitude of the overall envelope ($|\tilde{R}|=\sqrt{T_c^2+T_s^2}$) follows the well-known Rician distribution. Appendix C further proves \eqref{naka_m} is the equivalent Nakagami's m-distribution.

\end{proof}

\end{theorem}

\begin{remark}
The pre-conditions of \textbf{Theorem \ref{tt2}} hold true for a stationary receiver. If the receiver has a non-negligible speed towards or away from the RIS, the Doppler shift will induce a non-zero covariance between \eqref{tc} and \eqref{ts}. In those cases, the Rician distribution is not a good approximation for the joint fading channel. 
\end{remark}
\begin{remark}
It is worth mentioning that the asymptotic behaviour differs for these two distribution models. For Rician fading, the slope of the outage probability versus SNR is the same as for Rayleigh fading. For Nakagami fading, the slope is steeper, similar to that of m-branch diversity reception of a Rayleigh fading signal. As a result, using Nakagami distribution with $m = \frac{K^2+2K^{Eff}+1}{2K^{Eff}+1}$ to analyze the outage probability in the high SNR region leads to overly optimistic results.
\end{remark}

\section{Channel Model for Anomalous Reflecting RIS}\label{case}
In this section, we consider the case in which the RIS function as an anomalous reflector. To steer a plane-wave signal, the optimal configuration of the RIS is given by the co-phase condition~\cite{huang2008reflectarray}. This implies that the terms $\Phi(m,n)$ in \eqref{r_ris} are chosen to maximize $|\tilde{R}^{RIS}|$ at $\theta_{out}=\theta_{target}$ and $\varphi_{out}=\phi_{target}$. Under this setting, we discuss continuous phase shift and discrete phase shift, both deriving close-form expressions for the joint channel distribution.
\subsection{Continuous Phase shift}
\begin{figure}[t!]
    \begin{center}
        \includegraphics[width=3.0in]{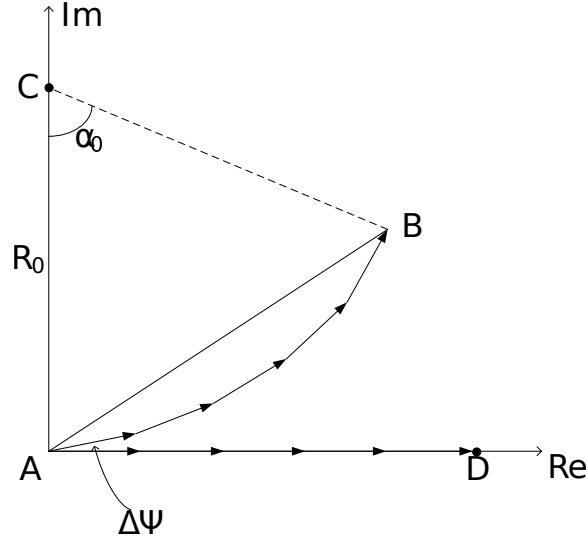}
        \caption{Illustration of the vector graph analysis}
        \label{vector}
    \end{center}
\end{figure}

First, we start from the continuous phase shift case which can be treated as the ideal limit of the discrete phase shift case. In this case, the phase shifts $\phi_{m}$ in \eqref{r_ris} can take on any value within $[0,2\pi)$. For convenience, we denote $\Phi_m=\phi_m-\epsilon m$, where $\epsilon = 2\pi \frac{p_x}{\lambda_c}(\sin\theta_{out}-\sin\theta_{in})$. As the angel of the receiver ($\theta_{out}$) varies, the additional phase shift associated with m-th element at $\theta_{out}$, namely $\Phi_m(\theta_{out})$, also changes. Implementing the co-phase condition for phase shifts, a continuous phase shift RIS is able to perfectly align all additional phase shifts at $\theta_{out}=\theta_{target}$. This can be expressed as:
\begin{equation}\label{align}
\Phi_m(\theta_{target}) = 0, \ \forall m \in (0,M].
\end{equation}
This is achieved by configure the phase shift of the RIS elements according to:
\begin{equation}\label{shift}
\phi_m = \frac{p_x}{\lambda_c}(\sin\theta_{target}-\sin\theta_{in})\cdot 2\pi m,
\end{equation}
It is also desirable to obtain the additional phase shift at directions other than the targeted one:
\begin{equation}\label{shift_other}
\Phi_m(\theta_{out})= \frac{p_x}{\lambda_c}(\sin\theta_{target}-\sin\theta_{out})\cdot 2\pi m,
\end{equation}
The above analysis implies that the time-irrelevant phase shifts of $\tilde{R}^{RIS}$ in \eqref{r_ris} are deterministic. On the contrary, the phase shift of $\tilde{R}^{D}$ in \eqref{r_d}, namely $\theta_n$, are randomly distributed over $[-\pi, \pi]$ because of their independent path length delays ($L_n/c\cdot 2\pi f_{D,n}$). As a result, when $N>>1$, meaning the number of the multipath components in the direct link is sufficiently large, $\tilde{R}^{D}$ exhibits a complex Gaussian distribution with zero mean. However, the deterministic $\tilde{R}^{RIS}$ provides a non-zero mean. The magnitude of this specular link can be approximated in the following fashion: As shown in Fig.~\ref{vector}, each arrow segment represents a term in summation \eqref{r_ris}. Since all $c_m$s are the same, the segments are of same lengths. The vector of length $\overline{AD}=M\cdot c_0$ represent $\tilde{R}^{RIS}$ at $\theta_{out}=\theta_{target}$. As the observation angle deviates from $\theta_{target}$, the arrow segments starts to bend, each of them makes a small tern of $\Delta\Phi=(p_x/\lambda_c)\cdot (\sin\theta_{out}-\sin\theta_{target})\cdot 2\pi$. As a result, these $M$ segments form a circular arc, approximately. Under the assumption of $M>>1$, and $\Delta\Phi<<1$, the length of the arc is equal to the length of $AD$. And thus, we have:
\begin{equation}\label{radius}
\overline{AB}=2R_0\sin(\frac{M\Delta\Phi}{2})=M\cdot c_0\cdot sinc(M\Delta\Phi/2).
\end{equation}
This result implies that the magnitude of $\tilde{R}^{RIS}$ at other observe directions is reduced by a factor of $sinc(M\Delta\Phi/2)$, compared with that at $\theta_{target}$.
The overall complexed envelope of the joint channel $\tilde{R}(t)$ can then be treated as a complex Gaussian distribution with a non-zero mean. It is well known that in this case, the signal envelope length $|\tilde{R}(t)|$ has a Rician distribution:
\begin{equation}\label{rice}
  p_{|\tilde{R}(t)|}(x)=\frac{2x(K+1)}{\Omega_p}e^{-K-\frac{(K+1)x^2}{\Omega_p}}I_0(2x\sqrt{\frac{K(K+1)}{\Omega_p}}),
\end{equation}
where the shape factor and scale factor are:
\begin{align}\label{k_conti}
&K=\frac{M^2\cdot c^2_0\cdot sinc^2(M\Delta\Phi/2)}{N\cdot \mathbb{E}[b_n^2]},\\
&\Omega_p=M^2\cdot c^2_0\cdot sinc^2(M\Delta\Phi/2)+N\cdot \mathbb{E}[b_n^2]
\label{scale_conti}
\end{align}

\begin{remark}
According to \eqref{cmn}, it can be proved that the amplitude of each multipath component ($c_0$) contains a Fraunhofer factor of $sinc(ku'p_x/2)$ induced by the diffraction at each RIS element. And according to \eqref{radius}, the overall amplitude of the received envelope contains a factor of $sinc(M\Delta\Phi/2)$ which can be regarded as another equivalent Fraunhofer factor on a larger scale, caused by the linear increase of the additional phase shifts along with the change in position of each RIS element.
\end{remark}

\subsection{Discrete Phase shift} \label{discrete}
In this case, $\phi(m,n)\in \{ t\cdot \frac{2\pi}{2^B} \}, t=0,1,2,...,2^B$ \cite{dunna2020scattermimo}. Let $\Delta = \frac{2\pi}{2^B}$, and $\theta^c_m = -\epsilon m+\phi_m$. \textbf{Lemma \ref{p2}} reveals how $\theta^c$ is distributed.

\begin{lemma}
\label{p2}
Consider the case when $M\epsilon >> \Delta$, which can be satisfied in practice for large $M$. Then, $\theta^c$ is uniformly distributed in the range: $(-\Delta/2,\Delta/2)$.
\begin{proof}
\textit{See Appendix D.}
\end{proof}
\end{lemma}
\begin{remark}
For the special case where $\epsilon=0$, meaning $\theta_{out}=\theta_{in}$. This case corresponds to a normal reflection behavior. The proposed model can be applied by setting $\Delta=0$ since no additional phase shifts are needed in this case.
\end{remark}

According to \textbf{Theorem \ref{tt1}, \ref{tt2}}, we have the following corollary that describes the distribution of the envelope in the case of discrete phase shift:
\begin{corollary}
\label{c2}
Suppose the expected magnitude of the each RIS multipath component and the direct link multipath component are $\Omega_r=\mathbb{E}[c^2_m]$ and $\Omega_d=\mathbb{E}[b^2_n]$, we denote the power ratio of these two as $K_0={M\Omega_r}/{(N\Omega_d)}$. At the target direction ($\theta_{out}=\theta_{target}$), the overall received envelope derived in \textbf{Corollary \ref{c1}} has a Rician distribution: $R(t) \sim \mathcal{R}(K^{Eff},\Omega_p)$, with shape factor and scale factor shown as follows:

\begin{align} \label{shape}
&K^{Eff}=\frac{Msinc^2(\Delta/2)}{1-sinc^2(\Delta/2)+K_0^{-1}}, \\ 
&\Omega_p =\Omega_r[M+(M^2-M)sinc^2(\Delta/2)]+N \Omega_d.
\label{scale}
\end{align}

\begin{proof}
\textit{See Appendix E.}
\end{proof}

\end{corollary}
\begin{remark}
It is helpful to obtain insights by observing several important limits of \textbf{Corollary \ref{c2}}. For convenience, define $s$ to be $sinc^2(\Delta/2)=\sin^2(\Delta/2)/(\Delta/2)^2$. When $K_0 \to 0$, it means that the power of the link via the RIS is negligible compared with the direct link, then $K^{Eff}=K_0Ms \to 0$, the joint channel exhibit a Rayleigh fading. When $\Delta \to 0$ i.e. $s \to 1$, it indicates a continuous (perfect) phase shift scenario, then $K^{Eff} \to M\cdot K_0$ and $\Omega_p \propto M^2\Omega_r + N\Omega_d$. In this scenario, the shape factor scales linearly with $M$, and the scale factor (which is the average received envelope power) scales with $M^2$. This $M^2$ dependency is also derived in \cite{basar2019wireless} by adopting a crude two-ray system model.
\end{remark}

\begin{remark}
The power scaling law for the anomalous RIS link under the discrete phase shift case can be obtained from \eqref{scale}:
\begin{equation}\label{scaling_law}
P_r \approx P_t \cdot C \cdot\big[sinc^2\frac{\Delta}{2}M^2 + (1-sinc^2\frac{\Delta}{2})M\big]\cdot,
\end{equation}
where $P_t$ is the transmit power and $C$ is a constant. According to our model, the $M^2$ dependency appears when the RIS phase error can be ignored (phase terms of different half-channels are perfectly aligned). When the RIS performs no phase shifts, correspond to $\Delta = 2\pi$, the average received power scales linearly with the number of elements (M).
\end{remark}

\section{Performance Analysis: RIS with Multiple Access Scheme}
\label{analysis}
In this section, we study the case where multiple users are served by a transmitter with the assistance of one RIS. Suppose the $k$-th user is located at $\theta^{(k)}_{out}$ with respect to the RIS. According to \textbf{Theorem \ref{tt2}}, the joint channel between the transmitter and the user can be modelled as a Rician channel: $\tilde{h}_k(K_k,\Omega_{pk})$. For different multiple access scheme, the receiver SNR distribution for each user is derived. The outage probability and its asymptotic behaviour is also studied.

\subsection{NOMA}
Consider the use of NOMA on a group of $q$ users. The transmitter broadcasts a combination of messages to all NOMA users, and the observation at the $k$-th user is given by:
\begin{equation}\label{signal}
    y_k = \tilde{h}_k(K_k,\Omega_{pk})\sum_{p=1}^{q}\sqrt{a_pP_t}s_p+n_k,
\end{equation}
where $n_k$ is the additive white Gaussian noise at the $k$-th user. We assume that each $n_k$ is distributed with a variance of $\sigma^2$. $a_p$ is the power allocation coefficient for the $p$-th user, $s_p$ is the information for the $p$-th user. $\tilde{h}_k(K_k,\omega_{pk})$ is the joint channel between the transmitter and the receiver. Note that the joint channels for each of the users are correlated. According to \eqref{k_conti} and \eqref{scale_conti}, the shape factors and scale factors for these Rician distributions depends on the choice of the target angle ($\theta_{target}$). We denote $\Delta\Phi_k=(p_x/\lambda_c)\cdot(\sin\theta^{(k)}_{out}-\sin\theta_{target})\cdot 2\pi$. Then for the continuous phase shift case, we have:
\begin{align}\label{h1h2}
&K_k=\frac{M\cdot c_0\cdot sinc(M\Delta\Phi_k/2)}{N\cdot \mathbb{E}[b_n^2]},\\
&\Omega_{pk}=M\cdot c_0\cdot sinc(M\Delta\Phi_k/2)+N\cdot \mathbb{E}[b_n^2].
\label{h1h2_t}
\end{align}
As a result, different choice of target angle ($\theta_{target}$) will result in different joint channels for all NOMA users. For RIS configurations which are not governed by the co-phase condition, $\tilde{h}_k$ need to be further calculated according to \textbf{Theorem \ref{tt1}}.
Without loss of generality, we assume the joint channel for different user follow the order as: $|\tilde{h}_1|^2 \leq |\tilde{h}_2|^2 \leq ... \leq |\tilde{h}_q|^2$. The power allocation coefficients are assumed to follow the order as: $a_1 \geq a_2 \geq ...\geq a_q$.

Next, we consider the signal to interference plus noise ratio (SINR) for each user. Consider the $k$-th user, according to the NOMA principle, it need to decode the message of all user $p$ with $p<k$. The message of user $p$ with $p>k$ are treated as interference. As a result, the SINR for the $k$-th user to decode the information of the $l$-th user is given by
\begin{equation}\label{sinr_kl}
    \gamma_{k,l}=\frac{|\tilde{h}_k|^2P_ta_l}{|\tilde{h}_k|^2P_t\sum_{p=l+1}^q a_p+\sigma^2}
\end{equation}

\subsection{FDMA and TDMA}
Suppose that the transmitter assign two equal sized frequency bands that is orthogonal to each other, the SNR for each user can be expressed as:
\begin{equation}\label{fdma}
\gamma_i = \frac{|\tilde{h}_i|^2 P_i}{\sigma^2/2},
\end{equation}
where $P_i$ is the transmit power allocated to user $i$ in FDMA scheme.

In TDMA, for both users, we have:
\begin{equation}\label{tdma}
\gamma_i = \frac{|\tilde{h}_i|^2 P_{t}}{\sigma^2}.
\end{equation}

\subsection{Outage Probability}

The outage probability is defined as:
\begin{equation}\label{outage_define}
Pr_{out}=Pr\{\gamma< \gamma_{min}\},
\end{equation}
where $\gamma$ is the SNR experienced by the user, $\gamma_{min}$ is the target SNR value chose by the transmitter for a data rate $C=B log_2(1+\gamma_{min})$.
According to the definition, the outage probability can be rewritten as:
\begin{equation}\label{outage2}
  P_i = 1-Pr\{|\tilde{h_i}(t)|^2 > \mu_i\},
\end{equation}
where the $\mu_i$ depends on different multiple access schemes and the target SNR ($\gamma_{min}$). 

For FDMA and TDMA, we have:
\begin{equation}\label{fdma_mu}
\mu^{FDMA}_i = \frac{\gamma_{min}\sigma^2}{2P_i},\ \  \mu^{TDMA}_i = \frac{\gamma_{min}\sigma^2}{P_{t}}.
\end{equation}
where $i$ indicating different users. 
%%%%%%%%%%%%%%%%%%%%%%%%%%%%%%%%%%%%%%%%%%%%%%%%%%

For the NOMA users, the user $k$ will declare an outage if it cannot successfully decode messages for user $l$ with $l\leq k$:
\begin{equation}\label{u1_out}
P_k=1- Pr\{\gamma_{k,l}>\tau_l \}, \ \ \ \forall l\leq k.
\end{equation}
Comparing with \eqref{outage2} and using \eqref{sinr_kl}, we have:
\begin{equation}\label{mu_1}
\mu_k = \frac{\tau_l}{a_l-\tau_l\sum_{p=l+1}^q a_p}\cdot\frac{\sigma^2}{P_t}, \ \ \ \forall l\leq k.
\end{equation}
In other words, if we denote $\tau_k=\frac{\tau_k}{a_k-\tau_k\sum_{p=l+1}^q a_p}\cdot\frac{\sigma^2}{P_t}$, we have:
\begin{equation}\label{mu_2}
\mu_k = min\{\tau_1, \tau_2,..., \tau_k \}.
\end{equation}
After clarify the expression for $\mu_i$ for different multiple access scheme, the outage probabilities can be written out in a general form:
\begin{corollarybox}
\label{c_multi}
For user $i$, the expressions for the outage probability is given by:
\begin{align}\label{outage_2}
  &P_i = 1-Q_1(\sqrt{2K_i},\sqrt{2\mu_i(K_i+1)/\Omega_{pi}}),\\
\end{align}
where $K_i$ and $\Omega_{pi}$ are given by \eqref{h1h2} to \eqref{h1h2_t}, and $Q_1(a,b)$ is the Marcum Q-function, defined as:
\begin{equation}\label{mq}
  Q_1(a,b)=\int_{b}^{\infty}x \cdot exp(-\frac{x^2+a^2}{2})I_0(ax)dx.
\end{equation}
\end{corollarybox}

Next, the behaviour of the cumulative distribution of $R(t)$ for amplitudes near zero can be studied. It is well known that the c.d.f of the squared Rician distributed envelope follows the non-central chi-square distribution:
\begin{equation}\label{chi}
\gamma\sim\frac{K^{Eff}+1}{\overline{\gamma}} e^{ -K^{Eff}-\frac{(K^{Eff}+1)\gamma}{\overline{\gamma}}} I_0(\sqrt{\frac{K^{Eff}(K^{Eff}+1)\gamma}{\overline{\gamma}/4}}),
\end{equation}
where $\gamma=|R(t)|^2/N_0$ and $\overline{\gamma}=\Omega_p/N_0$. For a small threshold 
$\gamma_{th} << \overline{\gamma}$, the c.d.f. can be approximate as:
\begin{equation}\label{thresh}
Pr(\gamma < \gamma_{th}) \to (1+K^{Eff})e^{-K^{Eff}}\frac{\gamma_{th}}{\overline{\gamma}}+O(\gamma_{th}^2)\to O(\gamma_{th})
\end{equation}
This result shows that the effective shape factor $K^{Eff}$ affects the outage probability in the high SNR region. A higher $K^{Eff}$ means a smaller outage probability under the same scale factor and threshold $\gamma_{th}$. However, it does not affect the asymptotic behavior of the system. In other words, according to our model, the diversity order of the system is fixed to one.

\begin{remark}
Although in our proposed joint channel model, the diversity order does not increase with the number of elements of the RIS, we need to point out that a higher diversity order may be observed in other settings, for example, when the RIS links are considered as NLoS links, as in \cite{ding2020impact}. In those cases, the signal impinging on the RIS has a nearly random phase and exhibits a power distribution itself. This will increase the difficulties to configure the RIS. However, if the RIS can be appropriately configured, the system can obtain a higher diversity order than that in our proposed model.
\end{remark}

\section{Numerical Results}\label{num}
In this section, numerical results are presented to facilitate the performance evaluations of the RIS-assisted wireless network. We aim to confirm the effectiveness of the proposed theorem by comparing the analytical derivations with simulation results. After that, the predicted outage probabilities are tested against the simulation results.

\subsection{The Rician Distribution of the Received Envelope}

\begin{figure}[t!]
    \begin{center}    
        \includegraphics[width=3.5in]{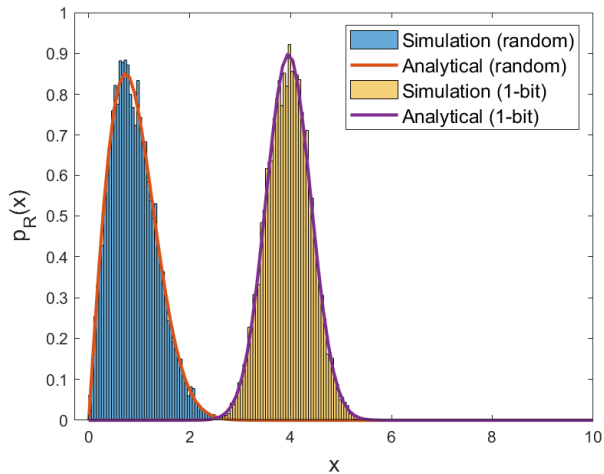}
        \caption{Simulated envelope distribution and Rician distribution (blue:random phase shifts, yellow: one-bit discrete phase shifts)}
        \label{r_pdf}
    \end{center}
\end{figure}

First, an intuitive test is implemented to show how well the RIS-aided joint channel fits into a Rician distribution, as proposed in \textbf{Corollary \ref{c2}}. The simulation is carried out in the following fashion: Firstly, each specular signal component is generated with a random phase under the assumption of \textbf{Lemma \ref{p2}}. Secondly, the received envelope is calculated by combining the inphase and quadrature components, as in \textbf{Corollary \ref{c1}}. Next, we take an iteration of 10000, in each of the iterations, we calculate a received signal envelope so that the collection of this variable is independent and identically distributed. Finally, we fit the histogram of the 10000 simulated envelopes to a Rician distribution and obtain the simulated shape factor and scale factor. As shown in Fig.~\ref{r_pdf}, the simulation runs under the condition where $M=50$ and $K_0=3$. The simulated distribution of the received envelope is well fitted to the analytical Rician distribution. Moreover, the effect of the one-bit discrete phase shift RIS link is compared with the random phase configuration. The fitted Rician distribution has a shape factor of $0.0186$ for the random phase shift and $39.1398$ for the one-bit discrete phase shift. 

\subsection{Effect of $K^0$ on $K^{Eff}$}

\begin{figure} [t!]
\centering
\includegraphics[scale=0.6]{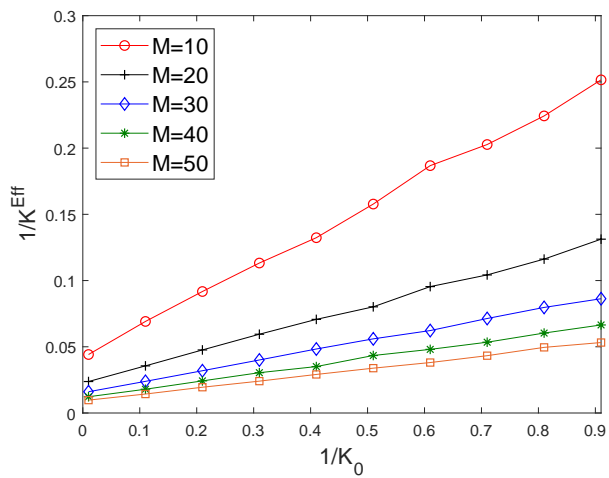}
 \caption{Relation between simulated $K^{Eff}$ and $K_0$}
 \label{k_k}
\end{figure}

The power ratio $K_0$ is defined as:
\begin{equation}
    K_0=\frac{M\mathbb{E}[c^2_m]}{N\mathbb{E}[b^2_n]},
\end{equation}
where $c_m$ represents the amplitude of the signal reflected through the m-th column on the RIS and $b_n$ represents the amplitude of the n-th multi-path signal. As shown in Fig.~\ref{k_k}, in this simulation, $K_0$ is varied in a range from 1 to 100, while other parameters in \eqref{shape} are fixed as: $\Delta = \pi$, i.e.($B=1$) and $M$ takes on 5 separate values.

From the simulation result, it can be observed that the inverse of the effective shape factor of the joint channel $K^{Eff}$ has a linear relationship with the inverse of $K_0$. This observation is consistent with the analytical result given in \textbf{Corollary \ref{c2}}, since the equation can be rewritten as:
\begin{equation}\label{k_eff_k}
    1/K^{Eff} = \frac{1}{M sinc^2(\Delta/2)}\cdot 1/K_0 + \frac{1-sinc^2(\Delta/2)}{M sinc^2(\Delta/2)}.
\end{equation}

Moreover, the slope and intercept distance with the $1/K^{Eff}$-axis of the curve decrease with the increase of $M$ as predicted in (\ref{k_eff_k}).

\subsection{Outage Probability for a Single User}
\begin{figure}[t!]
    \begin{center}
        \includegraphics[width=3.4in]{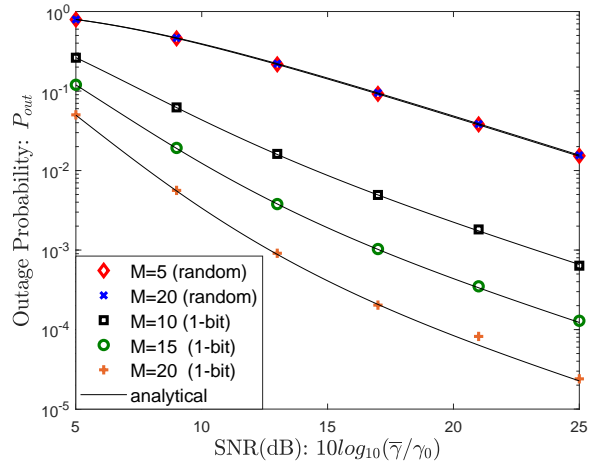}
        \caption{Simulated outage probability plotted against the analytical results (from top to bottom: random phase shift, 1-bit discrete phase shift with $M$= 10, 15, 20)}
        \label{out}
    \end{center}
\end{figure}

As shown in Fig.~\ref{out}, in this simulation, the transmitted signal's SNR is varied in a range from $5$dB to 25dB, while from top to bottom, the lines correspond to random phase shift, 1-bit discrete phase shifts with $M= 10, 15, 20$. The power ratio of $K_0$ is fixed to $1$. The Monte Carlo based simulation is carried out by generating $10^6$ independent and identically distributed signal envelopes. It can be observed that the analytical Marcum Q-function fits well with the simulation results. 

For the case of random phase shift, there is no observable diversity gain in the outage probability by increasing the number of elements of the RIS (if $K_0$ is fixed). This is predicted by \textbf{Corollary \ref{c2}} since when $\Delta = 2\pi$ and $K^{Eff}=0$, the overall channel exhibits a Rayleigh distribution. Simulation results verified this as the curve in the top of Fig.~\ref{out} is the result of two curves ($M=5$ (random) and $M=20$ (random)) coincide together. For the case of one-bit discrete phase shift, power saving can be achieved from the diversity gain by increasing the number of elements of the RIS. However, the increasing speed of this gain starts to diminish as the number of elements becomes larger. For example, going from $M=10$ to $M=15$, at $1\%$ outage probability there is an approximate $4$ dB reduction in the required SNR. However, from $M=15$ to $M=20$, the reduction in required SNR less than $2$ dB.

\subsection{Outage Probability for NOMA}
\begin{figure}[t!]
    \begin{center}
        \includegraphics[width=3.4in]{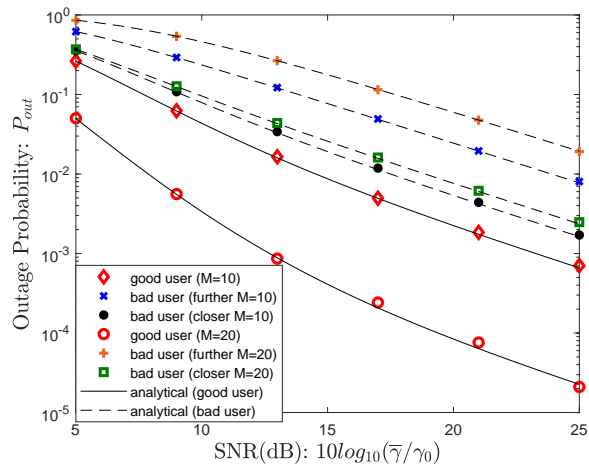}
        \caption{Simulated outage probability plotted against the analytical results, for the NOMA user pair}
        \label{out_noma}
    \end{center}
\end{figure}

As shown in Fig.~\ref{out_noma}, in this simulation, the transmitted signal's SNR is varied in a range from $5$dB to 25dB. We simulate the 1-bit discrete phase shifts beam steering scenario with $M=10$ and $20$ for two NOMA users. The RIS is configured to target the good user. Simulation results show that the good user experienced a smaller outage probability for both $M=10$ and $M=20$ cases. The outage probability of the good user decreases as $M$ increases, similar to the results in Fig.~\ref{out}. However, for bad users, its outage probability does not decrease monotonously with an increase of $M$, it further depends on the angle between the two users. For example, as shown in Fig.~\ref{out_noma}, when the bad user is further away from the good users (further away from the targeted angle), the outage probability is larger when $M=20$, compared with the one when $M=10$. Moreover, it can be observed that for both users, the slope of the curve reaches that same asymptotic limit in the high SNR region.

\subsection{Comparing MA schemes}
\begin{figure}[t!]
    \begin{center}
        \includegraphics[width=3.4in]{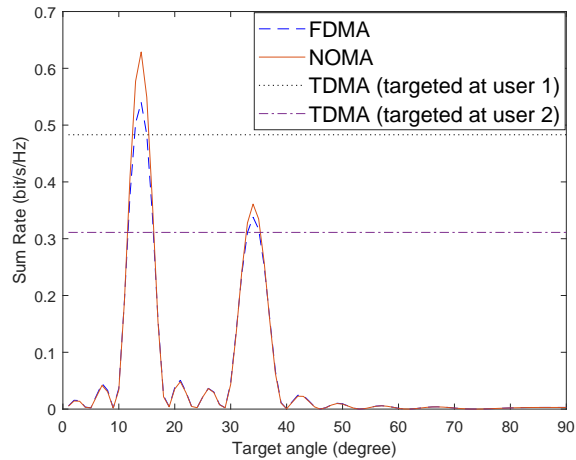}
        \caption{Simulated sum rate v.s. RIS target angle for different MA schemes (user 1 is located at $15^\circ$ w.r.t the RIS, and user 2 is located at $33^\circ$, the power allocation for the good user and the bad user are set to be $0.4$ and $0.6$, respectively).}
        \label{target_angle}
    \end{center}
\end{figure}
Based on our proposed model, the performance of the RIS-assisted channel, under different phase shift configurations, can be studied for different MA schemes: As illustrated in Fig.~\ref{target_angle}, the best sum rate can be achieved by adopting NOMA and configure the RIS to target user 1. When the target angle is configured to be near $15^\circ$ or $33^\circ$, NOMA shows superiority compared to both FDMA and TDMA, in terms of sum rate. As the target angle moves from user 1 to user 2, the channel condition of user 1 decreases, while the channel condition for user 2 increases. A possible switch could occur between the good user and the bad user. Moreover, one can obtain insight from Fig.~\ref{target_angle} that the best sum rate is achieved when the RIS is configured to target the user with the best achievable channel gain.
\section{Conclusions}\label{conclusion}
%\vspace{-0.5cm}
A compact joint channel model for the RIS-assisted wireless communication network was proposed. The statistical multipath analysis and the physics-based radiation pattern calculation were reconciled. Analytical results for the joint channel distribution were derived. We demonstrated that the joint channel exhibit a Rician distribution, where the shape of the distribution depends on the number of elements of the RIS, the quantization level of phase shift, the observing angle of the receiver, and the power ratio between the specular link through the RIS and the direct link. The validity of our proposed models depends on the assumptions of a flat fading channel and the vertically polarized signal. Moreover, the links between the BS and the RIS and between the RIS to the receiver were assumed to be LoS-dominant links. Different models for the joint channel in other RIS application scenarios remains an open question.
\section*{Appendix~A: Proof of \textbf{Theorem \ref{tt1}}} \label{Appendix:A}
The focus of this proof is on \eqref{r_ris}, which indicates that the specular link through the RIS can be written as the 2-D discrete Fourier transform of the phase shift configuration matrix. First, we write down the received signal(the reflected electric field through RIS) in terms of the incident field, the incident angle, and the reflected angle. As shown in Fig. \ref{sys}, consider the frame of reference where the RIS lays in $z=0$ plane, and both the BS and terminal are in $y=0$ plane. The tangential incident electric field can be expressed in Cartesian coordinates as:
\begin{equation}\label{incident}
\bm{E}_{1}(x,y) = E_{1x}(x,y)\hat{x} + E_{1y}(x,y)\hat{y}.
\end{equation}

Suppose after reflected by the RIS, the reflected tangential field becomes:
\begin{equation}\label{e_out_1}
  \bm{E}_2(x,y)=\tilde{r}(x,y) \cdot \bm{E}_1(x,y)=E_{2x}(x,y)\hat{x}+E_{2y}(x,y)\hat{y}.
\end{equation}
Using the stationary-phase approximation \cite{nayeri2018reflectarray}, we could write down the far field radiation pattern as:
\begin{align}\label{book}
\begin{split}
  E(\theta,\varphi)=jk[(\hat{\theta}cos\varphi-\hat{\varphi}sin\varphi cos\theta)\tilde{E}_{2x}(\theta,\varphi)+ \\
  (\hat{\theta}sin\varphi-\hat{\varphi}cos\varphi cos\theta)\tilde{E}_{2y}(\theta,\varphi)]\frac{e^{-jkr_2}}{2\pi r_2},
\end{split}
\end{align}

where $\tilde{E}_{2y}(\theta,\varphi)$ are defined as
\begin{align}\label{e_tiled}
\begin{split}
  \tilde{E}_{2x/y}(\theta,\varphi)=&\iint_{\Sigma_{RRS}} E_{2x/y}(x,y)exp[jk(\sin\theta \cos\varphi \cdot x+ \\
  &\sin\theta \sin\varphi \cdot y)]dxdy.
\end{split}
\end{align}

For simplicity, we set $u=\sin\theta \cos\varphi$ and $v=\sin\theta \sin\varphi$.   
In order to transform integral into a summation over each element on the RIS, we have to evaluate the integral in \eqref{e_tiled} element by element. For the $(m,n)$-th element on the RRS, the range of the coordinates are
\begin{align}\label{xy}
\begin{split}
  x & =x'+mp_x-\frac{(M_x-1)p_x}{2}; m=0,1,...,M_x-1, \\
  y & =y'+np_y-\frac{(M_y-1)p_y}{2}; n=0,1,...,M_y-1,
\end{split}
\end{align}
where $M_x$ and $M_y$ are the number of elements along each directions.
By substituting \eqref{xy} into \eqref{e_tiled}, the spectral functions for the $x/y$ components are written as:
\begin{align}\label{components}
\begin{split}
\tilde{E}_{2x/y}(u,v)=&K_1\sum_{m=0}^{M_x-1}\sum_{n=0}^{M_y-1}\bigg[ e^{jk_0(ump_x+vnp_y)} \\
 & \cdot\int_{-p_x/2}^{p_x/2}\int_{-p_y/2}^{p_y/2}E^{mn}_{2x/y}(x',y')e^{jk_0(ux'+vy')}dx'dy' \bigg],
\end{split}
\end{align}
where
\begin{equation}\label{k1}
K_1=e^{-j\frac{k_0}{2}[u(M_x-1)p_x+v(M_y-1)p_y]},
\end{equation}
and $E^{mn}_{2x/y}(x',y')$ is uniform in each cell, its value depends on the amplitude response ($r(m,n)$) and phase-shift response ($\phi(m,n)$) of the $(m,n)$-th cell:
\begin{equation}\label{mn}
E^{mn}_{2x/y}(x',y')=r(m,n)e^{j\phi(m,n)}.
\end{equation}
Since $E^{mn}_{2x/y}(x',y')$ does not depend on the resized positions $(x',y')$, the double integral in \eqref{components} can be easily calculated:
\begin{align}\label{longlong}
\begin{split}
\tilde{E}_{2x/y}(u,v)=&K_1p_xp_y sinc(\frac{k_0up_x}{2})sinc(\frac{k_0vp_y}{2})\\
&\cdot \sum_{m=0}^{N_x-1}\sum_{n=0}^{N_y-1}r(m,n)e^{j\phi(m,n)}e^{jk_0(ump_x+vnp_y)}.
\end{split}
\end{align}
The double summation in \eqref{longlong} can be written as a 2-D inverse discrete Fourier Transform (IDFT2), which is defined as:
\begin{equation}\label{idft2}
f(p,q)= IDFT2[F(m,n)]=\frac{1}{MN}\sum_{m=0}^{M-1}\sum_{n=0}^{N-1}F(m,n)e^{j\frac{2\pi m}{M}p}e^{j\frac{2\pi n}{N}q}.
\end{equation}
By comparing \eqref{idft2} and \eqref{longlong}, it is clear that the variables $(p,q)$ is related to the direction indicators $(u,v)$ in the following fashion:
\begin{align}\label{pquv}
&u=\frac{2\pi}{M_xp_xk_0}p;  \ \ p=0,1,2,...M_x-1,\\
&v=\frac{2\pi}{M_yp_yk_0}q;  \ \ q=0,1,2,...M_y-1.
\end{align}

As a result, by plugging \eqref{longlong} back into \eqref{e_tiled}, and absorbing coefficients outside the IDFT2 into $c(m,n)$, we can reach the same form as \eqref{r_ris} for both $\tilde{E}_{2x}(\theta,\varphi)$ and $\tilde{E}_{2y}(\theta,\varphi)$.

\section*{Appendix~B: Proof of \textbf{Corollary \ref{c1}}} \label{Appendix:B}

Since we assumed that the signal’s field is vertically polarized(along $y$ direction), the tangential field along two directions at the RIS can be written as:

\begin{equation}\label{e1x}
  E_{1x}(x,y)= 0,
\end{equation}
\begin{equation}\label{e1y}
  E_{1y}(x,y)= E_0 \cdot e^{-jkx\cos\alpha} = E_0 \cdot e^{-jkx\sin\theta_{in}}.
\end{equation}
After reflected by the RIS, the reflected tangential field becomes:
\begin{equation}\label{e_out_2}
  E_2(x,y)=\tilde{r}(x,y) \cdot E_1(x,y) = E_{2y}(x,y)\hat{y}.
\end{equation}
As a result, the integral in \eqref{e_tiled} evaluated in the area within the $(m,n)$-th element on the RRS should be:
\begin{equation}\label{emn1}
  \tilde{E}^{mn}_{2y}=K\int_{-p_x/2}^{p_x/2}\int_{-p_y/2}^{p_y/2}E^{mn}_{2y}exp[jk(ux'+vy')]dx'dy'.
\end{equation}
Since $E^{mn}_{2y}=\tilde{r}(m,n)E^{mn}_{1y}= E_0e^{-jkx\sin\theta_{in}}e^{j\phi(m,n)}$. Since in the system model, the terminal(receiver) is in $y=0$ plane, we have $\varphi = 0$. As a result, let $u'=u-\sin\theta_{in}$,we have:
\begin{align}\label{emn2}
  \tilde{E}^{mn}_{2y}=K_1|E_0| e^{j\phi(m,n)} e^{jk(u'mp_x)}p_xp_ysinc(\frac{ku'p_x}{2}),
\end{align}
where $K_1=exp\{-j\frac{k}{2}(u'(M_x-1)p_x)\}$. Next, collecting the contribution of the $(m,n)$-th element to the $\hat{\varphi}$ direction of the reflected field is:
\begin{align}\label{phi_out}
  E^{mn}_{\varphi}(\theta_{out}) 
     = |E^{mn}_{\varphi}(\theta)| e^{-j\Phi_{mn}},
\end{align}
where:
\begin{align}\label{cmn}
    &|E^{mn}_{\varphi}(\theta_{out})|=E_0
     \frac{k}{4\pi d_1d_2} p_xp_y sinc(\frac{ku'p_x}{2})\cos\theta_{out}, \\
     &\Phi_{mn} = ku'(M_x-1)\frac{p_x}{2}+k(d_1+d_2)+\phi(m,n)-ku'mp_x).
\end{align}

Finally, we can arrive at the received signal:
\begin{align}\label{sum_re}
  r(t) &= Re[\sum_{m=0}^{M_x-1} \sum_{n=0}^{M_y-1}|E^{mn}_{\varphi}(\theta)| e^{-j\Phi_{mn}}e^{jw_ct}] \\
  &=T^{RIS}_c(t)\cdot \cos(\omega_ct)+T^{RIS}_s(t)\cdot \sin(\omega_ct).
\end{align}

As a result, if we let $\theta_0= ku'(M-1)p_x/2+k(d_1+d_2)$, $\epsilon = ku'p_x$ and $c(m,n)==|E^{mn}_{\varphi}(\theta_{out})|$, we have:
\begin{align}\label{tc_ris_proved}
  &T^{RIS}_c(t) = \sum_{m=1}^{M_x}\sum_{n=1}^{M_y}c(m,n)\cos(\omega_0t+\theta_0-\epsilon m+\phi(m,n)), \\
  &T^{RIS}_s(t) = \sum_{m=1}^{M_x}\sum_{n=1}^{M_y}c(m,n)\sin(\omega_0t+\theta_0-\epsilon m+\phi(m,n)).
\end{align}
When the amplitude response ($c(m,n)$) and the phase-shift response ($\phi(m,n)$) do not depend on the row index $n$, we have the inphase and quadrature components as expressed in \textbf{Corollary \ref{c1}}.

\section*{Appendix~C: Proof of \textbf{Theorem \ref{tt2}}} \label{Appendix:C}
According to \eqref{complex}, the magnitude of the overall received signal's is:
\begin{equation}\label{mag_signal}
|\tilde{R}(t)|=|T_c+jT_s|.
\end{equation}
Base on Nakagami's original derivations, when the condition of central limit theorem holds, $|\tilde{R}(t)|$ follows the $m$-distribution, and the parameter $m$ takes the form:
\begin{equation}\label{naka-m}
    m = \frac{(\sigma+A^2)^2}{(\sigma+A^2)^2+(B^2-A^4)+2A^2B\cos{2(\delta_1-\delta_2)}},
\end{equation}
where
\begin{align}\label{long}
    &A^2 = (\mathbb{E}[T_c])^2+(\mathbb{E}[T_s])^2, \\
    &\sigma_x = Var(T_c) ,\ \sigma_y = Var(T_s),\\
    &c = cov(T_c,T_s),\\
    &B^2  = 4c^2+(\sigma_x-\sigma_y)^2,\\
    &\sigma = \sigma_x+\sigma_y ,\\
    &\delta_1 = \arctan{\frac{\mathbb{E}[T_s]}{\mathbb{E}[T_c]}},\\
    &\delta_2 = \frac{1}{2}\arctan\frac{2c}{\sigma_x-\sigma_y}.
\end{align}
The pre-conditions of \textbf{Theorem \ref{tt2}} indicates that $B^2=0$ and thus, the parameter $m$ is the same as shown in \eqref{naka_m}.

\section*{Appendix~D: Proof of \textbf{Lemma \ref{p2}}} 
\label{Appendix:D}
Consider the following sketch of $\epsilon m-\phi_m$ v.s. $\epsilon m$. Since in the discrete phase adjustment case, $\phi_m$ can take on $2^B$ different value within the range of $[0,2\pi)$, and the choice of $\phi_m$ should minimize the difference between $\epsilon m$ and $\phi_m$. As a result, the absolute difference between the two will never exceed $\Delta/2$. In other words: $\epsilon m-\phi_m \in [-\Delta/2,\Delta/2]$. When the ratio of $\epsilon$ and $\Delta$ is not a rational number, the quantity $\epsilon m-\phi_m$ will appear randomly in the range $[-\Delta/2,\Delta/2]$ for any given $m$.
\begin{figure} [ht!]
\centering
\includegraphics[scale=0.6]{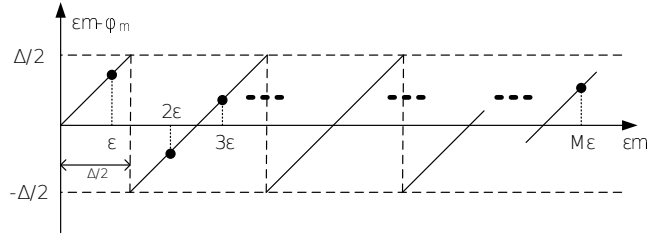}
 \caption{Sketch of $\epsilon m-\phi_m$ v.s. $\epsilon m$}
 \label{phi}
\end{figure}

\section*{Appendix~E: Proof of \textbf{Corollary \ref{c2}}} \label{Appendix:E}
The overall quadrature components in the case of discrete phase adjustment can be written as:
\begin{align}\label{tstc}
  T_c =& \sum_{m}^{M}c_m\cdot \cos\theta^c_m+\sum_{n}^{N}b_n\cdot \cos\theta^b_n, \\
  T_s =& \sum_{m}^{M}c_m\cdot \sin\theta^c_m+\sum_{n}^{N}b_n\cdot \sin\theta^b_n,
\end{align}
where $\theta^c$ is uniformly distributed in $[-\Delta/2,\Delta/2]$ and $\theta^b$ is uniformly distributed in $[0,2\pi]$, as proved in \textbf{Lemma \ref{p2}}. When $M,N>>1$, meaning the number of the elements on the RIS and the number of multi-paths is sufficiently large, according to central limit theorem, both $T_c$ and $T_s$ have a Gaussian distribution. As a result, they can be characterized by their means and variances. Moreover, the complex envelope $R = T_c + jT_s$, we have:
\begin{align}\label{mean_r}
  \mathbb{E}|R|] &= \mathbb{E}[T_c]=\mathbb{E} [\sum_{m=1}^{M}c_m\cdot \cos\theta^c_m + \sum_{n=1}^{N}b_n\cdot \cos\theta^b_n ] \\
   & =M\cdot \mathbb{E}[c_m]\cdot \int_{-\Delta/2}^{\Delta/2}\cos x\frac{1}{\Delta}dx \\
   &= M\cdot \mathbb{E}[c_m]\cdot sinc(\Delta/2),
\end{align}
and
\begin{align}\label{r2}
  \mathbb{E}[|R|^2] & = \mathbb{E}[T^2_c]+\mathbb{E}[T^2_s] \\
   & = \mathbb{E}[c^2_m][M+(M^2-M)sinc^2(\Delta/2)]+N\cdot \mathbb{E}[b^2_n].
\end{align}
If we assume the RIS only performs phase adjustment without any amplitude change, meaning $E^2[c_m]=E[c^2_m]$. As a result, we have:
\begin{align}\label{var_r}
  Var(R) & =\mathbb{E}[|R|^2]-\mathbb{E}^2[|R|] \\
   & =\mathbb{E}[c^2_m]M(1-sinc^2(\Delta/2))+N\cdot \mathbb{E}[b^2_n].
\end{align}
Thus, we can obtain the effective shape factor $K^{Eff}$ and the scale factor $\Omega_p$:
\begin{align}\label{eff}
  &K^{Eff} =\frac{\mathbb{E}^2[|R|]}{Var(R)}=\frac{Msinc^2(\Delta/2)}{1-sinc^2(\Delta/2)+K^{-1}_0}, \\
  &\Omega_p =\mathbb{E}[c^2_m][M+(M^2-M)sinc^2(\Delta/2)]+N\cdot \mathbb{E}[b^2_n].
\end{align}

%\vspace{-0.4in}
\linespread{1.2}

\bibliographystyle{IEEEtran}
\bibliography{mybib}

\end{document}